\documentclass[reprint,
superscriptaddress,amsmath,amssymb,aps, longbibliography
]{revtex4-2}

\usepackage{graphicx}% Include figure files
\usepackage{dcolumn}% Align table columns on decimal point
\usepackage{bm}% bold math
\usepackage{siunitx}
\usepackage[noend]{algorithmic}
\usepackage{algorithm}
\usepackage[skins]{tcolorbox}%colorbox
\usepackage{setspace}
\usepackage{blkarray}

\usepackage{xcolor}
\usepackage{soul}
\usepackage{systeme}
\usepackage{footmisc}
\usepackage{mhchem}

\usepackage{scalerel} % To change size things in math environment \scaleto{...}{1pt}

\makeatletter
\def\maketag@@@#1{\hbox{\m@th\normalfont\normalsize#1}}
\makeatother

\usepackage{booktabs}
\usepackage{hyperref}
\usepackage[capitalize]{cleveref}
\usepackage[utf8]{inputenc}

\usepackage[lining,semibold]{libertine} % a bit 
\usepackage{amsthm}
\usepackage[libertine, cmintegrals, bigdelims, vvarbb]{newtxmath}

\usepackage{amsmath}
\usepackage{amsfonts}
\usepackage{mathrsfs}
\usepackage{gensymb}
\usepackage{bbm}
\usepackage{dsfont}

\usepackage{pgfplots}
\usepgfplotslibrary{colormaps}

\usepackage{kbordermatrix}
\renewcommand{\kbldelim}{(}
\renewcommand{\kbrdelim}{)}

\usepackage{chemformula}
\usepackage[caption=false]{subfig}

\usepackage{soul}
\usepackage{xcolor}

\theoremstyle{definition}

\usepackage{scalerel} % To change size things in math environment \scaleto{...}{1pt}

\definecolor{webgreen}{rgb}{0,.5,0}
\definecolor{webbrown}{rgb}{.6,0,0}
\definecolor{grigio}{rgb}{.85,.85,.85} 
\definecolor{RoyalBlue}{rgb}{0.0, 0.14, 0.4}
\definecolor{skyblue1}{rgb}{0.45,0.62,0.81}
\definecolor{skyblue2}{rgb}{0.2,0.39,0.64}
\definecolor{skyblue3}{rgb}{0.13,0.29,0.53}
\definecolor{scarlet1}{rgb}{0.93,0.16,0.16}
\definecolor{scarlet2}{rgb}{0.8,0,0}
\definecolor{scarlet3}{rgb}{0.64,0,0}

\definecolor{g}{gray}{0.50}

\hypersetup{%
    colorlinks=true, linktocpage=true, pdfstartpage=1, pdfstartview=FitV,%
    breaklinks=true, pdfpagemode=UseNone, pageanchor=true, pdfpagemode=UseOutlines,%
    plainpages=false, bookmarksnumbered, bookmarksopen=true, bookmarksopenlevel=1,%
    hypertexnames=true, pdfhighlight=/O,%nesting=true,%frenchlinks,%
    urlcolor=webbrown, linkcolor=RoyalBlue, citecolor=webgreen, %pagecolor=RoyalBlue,%
    pdftitle={},%
    pdfauthor={Timur Aslyamov},%
    pdfsubject={},%
    pdfkeywords={},%
    pdfcreator={pdfLaTeX},%
    pdfproducer={LaTeX REVTeX}%
}

\begin{document}
%%%%%%%%%%%%%%%%%%%%%%%%%%%%%%%%%%%%% frontmatter %%%%%%%%%%%%%%%%%%%%%%%%%%%%%%%%%%%%%
\title{Macroscopic fluctuation-response theory and its use for gene regulatory networks}
%\title{Physical responses shape macroscopic fluctuations}
\author{Timur Aslyamov}
\email{timur.aslyamov@uni.lu}
\affiliation{Complex Systems and Statistical Mechanics, Department of Physics and Materials Science, University of Luxembourg, 30 Avenue des Hauts-Fourneaux, L-4362 Esch-sur-Alzette, Luxembourg}

\author{Krzysztof Ptaszy\'{n}ski}
\email{krzysztof.ptaszynski@ifmpan.poznan.pl}
\affiliation{Institute of Molecular Physics, Polish Academy of Sciences, Mariana Smoluchowskiego 17, 60-179 Pozna\'{n}, Poland}

\author{Massimiliano Esposito}
\email{massimiliano.esposito@uni.lu}
\affiliation{Complex Systems and Statistical Mechanics, Department of Physics and Materials Science, University of Luxembourg, 30 Avenue des Hauts-Fourneaux, L-4362 Esch-sur-Alzette, Luxembourg}

\date{\today}

\begin{abstract}
Gaussian macroscopic fluctuation theory underpins the understanding of noise in a broad class of nonequilibrium systems. We derive exact fluctuation-response relations linking the power spectral density of stationary fluctuations to the linear response of stable nonequilibrium steady states. Both of these can be determined experimentally and used to reconstruct the kernel of the linearized dynamics and the diffusion matrix, and thus any features of the Gaussian theory. We apply our theory to gene regulatory networks with negative feedback, and derive an explicit internal-external noise decomposition of the power spectral density for any networks, including cross-correlations.
\end{abstract}
\maketitle

\textit{Introduction---}The behavior of many nonequilibrium systems can be modeled as small stochastic fluctuations around their deterministic dynamics. This type of behavior is most commonly described using linear Langevin equations, a framework that is ubiquitous across physics and applied mathematics~\cite{elf2003fast,paulsson2005models,paulsson2004summing,lestas2010fundamental,yan2019kinetic,bruggeman2009noise,volpe2006torque,alonso2007stochastic,benayoun2010avalanches,bressloff2010stochastic,melbinger2012microtubule,szavitz2014inherent,dinis2012fluctuation,han2021fluctuation,aifer2024thermodynamic,gilson2023entropy,nicoletti2024tuning,fyodorov2025nonorthogonal,melanson2025thermodynamic}.
It can be derived in two complementary ways: either by linearizing the deterministic drift in a stochastic differential equation with additive Gaussian noise, an approach emphasized by van Kampen in his system-size expansion \cite{van1992stochastic,kubo1966fluctuation,gardiner2004handbook,risken1989fokker,stratonovich2012nonlinear}, or by considering the macroscopic limit of an underlying Markov jump process and expanding around its most probable deterministic trajectory \cite{kampen1961power,lax1960fluctuations,keizer2012statistical,kurtz1971limit}. In both perspectives, the resulting dynamics of fluctuations is Gaussian and governed by an Ornstein--Uhlenbeck process, representing the universal description of small deviations near stable deterministic states. This regime may be viewed as a Gaussian macroscopic fluctuation theory, providing the lowest order approximation in the noise intensity of the general nonlinear theory of fluctuations around nonequilibrium steady states~\cite{bertini2015macroscopic,falasco2023macroscopic}.

Stationary fluctuations are characterized by two-point correlation functions, or their Fourier transform, the frequency-dependent Power Spectral Density (PSD)[\cref{eq:PSD}]. When integrated over all frequencies, the PSD reduces to the stationary covariance. While devoid of dynamical content, it quantifies the overall intensity of stationary fluctuations and satisfies the Lyapunov equation [\cref{eq:Lyapunov}]. 
For systems obeying detailed balance (reciprocal dynamics), the fluctuation dissipation theorem~\cite{kubo2012statistical} provides an explicit relation for the stationary covariance in terms of the diffusion matrix and the static response of the system to perturbations [see \cref{eq:Kubo} with $\mathbb{Q}=1$].
However, far from equilibrium, the lack of time-reversibility breaks this connection~\cite{agarwal1972fluctuation,marconi2008fluctuation,seifert2010fluctuation,prost2009generalized,altaner2016fluctuation,chun2023trade,shiraishi2023introduction,gao2024thermodynamic,baiesi2009fluctuations,baiesi2013update,speck2016thermodynamic,maes2020response,tesser2024out,klinger2025computing} and a new nonequilibrium response theory is needed~\cite{lucarini2016response,santos2020response,falasco2019negative,mallory2020kinetic,owen2020universal,owen2023size,gabriela2023topologically,aslyamov2024nonequilibrium,aslyamov2024general,harunari2024mutual,cengio2025mutual,khodabandehlou2024affine,floyd2024learning,frezzato2024steady,floyd2024limits,gao2022thermodynamic,zheng2025spatial,auconi2025nonequilibrium,floyd2025local,ptaszynski2024dissipation,kwon2024fluctuation}.

In this letter, we show that another quantity derived from the PSD, namely its zero-frequency component, can always be expressed in terms of the system responses and diffusion matrix, even in the absence of detailed balance.
This relation is reminiscent of the Fluctuation-Response Relations (FRRs) recently discovered in the context of Markov jump processes \cite{aslyamov2024frr, ptaszynski2024frr, ptaszynski2025frr-mix, aslyamov2025excess} and of great practical value.
As an application, we consider models of gene regulatory networks describing stationary fluctuations in mRNA and protein (with negative feedback)~\cite{kaern2005stochasticity}. Standard approaches study stationary correlation functions and are built on the Lyapunov equation~\cite{paulsson2005models,paulsson2004summing,lestas2010fundamental,yan2019kinetic,bruggeman2009noise}. More recent approaches use the PSD of the auto-correlation to capture dynamical features of stationary fluctuations~\cite{simpson2003frequency,warren2006exact,gupta2022frequency,song2019frequency}. We provide an explicit expression for the full PSD matrix (including mRNA-protein cross-correlations) and show that it can be used to detect the presence of negative feedbacks. 
Most importantly, we show that our FRRs provide an unambiguous decomposition of the zero-frequency PSD in terms of intrinsic and extrinsic noise in arbitrary complex networks. This framework for decomposing noise draws inspiration from seminal experiments on gene expression variability~\cite{elowitz2002stochastic,pedraza2005noise}.

\begin{figure}
    \centering
    \includegraphics[width=\linewidth]{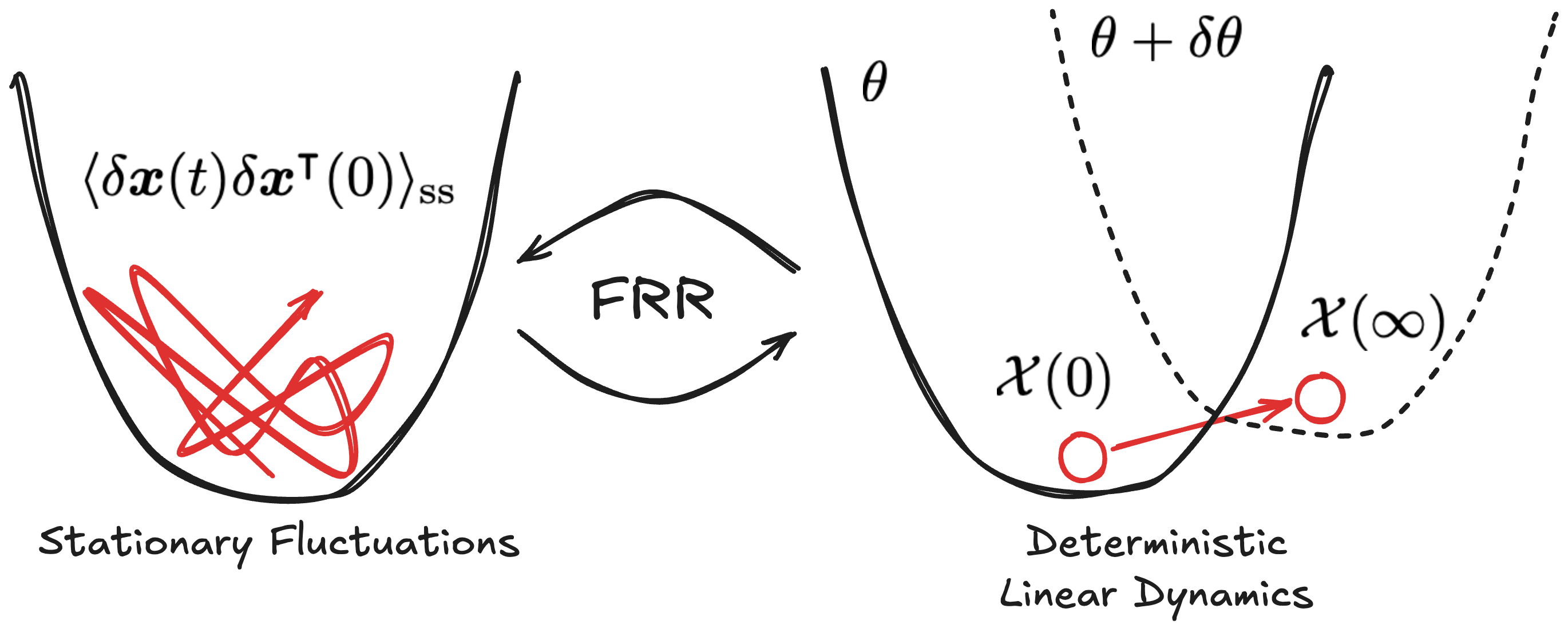}
    \caption{Linking the stationary fluctuations and responses far-from-equilibrium.}
    \label{fig:sketch}
    \vspace{-0.3cm}
\end{figure}

\textit{Macroscopic dynamics---}We consider a system described by the $N$-dimensional stochastic field $\boldsymbol{x}(t)$, which may correspond, e.g., to the particle position or a set of concentrations of chemical species. In the limit where the noise acting on the system becomes vanishingly small, the probability density of that field concentrates around the most likely value, $\mathcal{X}(t)$, which obeys the deterministic dynamical equation,
\begin{align}
\label{eq:dynamics-general}
d_t \boldsymbol{\mathcal{X}}(t) &= \boldsymbol{f}(\boldsymbol{\mathcal{X}}(t)) \,, \quad \boldsymbol{f}(\boldsymbol{x}^*)=0 \,,
\end{align}
where $\boldsymbol{f}(\boldsymbol{\mathcal{X}})$ is the rate vector and $\boldsymbol{x}^*$ is a fixed point which we assume to be stable and unique.  
At steady state and in presence of a small but non-vanishing noise, the fluctuations of the stochastic field $\boldsymbol{x}(t)$
in the vicinity of the fixed point are described by the linear Langevin equation
\begin{align}
\label{eq:Langevin-state}
    d_t\boldsymbol{x} = \mathbb{K} (\boldsymbol{x} -\boldsymbol{x}^*) + \sqrt{\varepsilon}\boldsymbol{\eta}\,.
\end{align}
Here, the first term describes the linear deterministic relaxation towards the fixed point, where 
\begin{align}
\mathbb{K} \equiv \partial_{\boldsymbol{x}} \boldsymbol{f}(\boldsymbol{x}) \vert_{\boldsymbol{x}=\boldsymbol{x}^*} 
\end{align}
is the Jacobian of the deterministic dynamics with the elements $K_{m k}=[\partial_{x_k} f_m(\boldsymbol{x})]_{\boldsymbol{x}=\boldsymbol{x}^*}$  for $\boldsymbol{f}(\boldsymbol{x})=(\dots, f_m(\boldsymbol{x}),\dots)^\intercal$ and $\boldsymbol{x}=(\dots, x_k, \dots)^\intercal$. The second term describes the effect of fluctuations as a Gaussian white noise vector $\boldsymbol{\eta}(t)=(\ldots, \eta_m (t), \ldots)^\intercal$ with zero average $\langle \boldsymbol{\eta} \rangle = 0$ and the autocorrelation function
\begin{align} \label{eq:noise-autocor-general}
\langle \boldsymbol{\eta}(t)  \boldsymbol{\eta}^\intercal(t') \rangle =\delta(t-t') \mathbb{D}\,,
\end{align}
where the average over noise realizations is denoted by $\langle\dots \rangle$ and $\mathbb{D}$ is the diffusion matrix that we assume to be positive semidefinite and constant. The prefactor $\varepsilon$ is the noise amplitude, which corresponds, e.g., to the temperature for colloidal particles or the inverse volume in chemical systems. 

We note that $\mathbb{K}$ and $\mathbb{D}$ are often not known. We will see in what follows that they can be determined from the measurable linear response of arbitrary steady states (beyond equilibrium) and their stationary fluctuations, see Fig. \ref{fig:sketch}.  

\textit{Stationary fluctuations---}Since \cref{eq:Langevin-state} is linear, the dynamics of the average of $\boldsymbol{x}$ is $d_t \langle \boldsymbol{x} \rangle=\mathbb{K} (\langle \boldsymbol{x} \rangle-\boldsymbol{x}^*)$, which is identical to the deterministic dynamics \cref{eq:dynamics-general} linearized around the fixed point.  
Its fluctuations are captured by the two-point correlation functions $\langle \delta \boldsymbol{x}(t+t') \delta \boldsymbol{x}^\intercal(t') \rangle$, where $\delta\boldsymbol{x}(t) = \boldsymbol{x}(t) - \langle\boldsymbol{x}(t)\rangle$. 
The \textit{dynamical covariance} is defined as $\mathbb{C}(t) \equiv \varepsilon^{-1} \langle \delta \boldsymbol{x}(t) \delta \boldsymbol{x}^\intercal(t) \rangle$.
At steady state, the mean and covariance become time invariant, $\langle\boldsymbol{x}(t)\rangle_\text{ss} = \boldsymbol{x}^*$ and $\mathbb{C}(t)=\mathbb{C}_\text{ss}$, and the correlation function becomes independent from $t'$ (which can thus be taken as $t'=0$). 
The \textit{power spectrum density} (PSD) of the steady state correlation function is  
\begin{align}
\label{eq:PSD}
    \mathbb{Z}(\omega) \equiv \frac{1}{\varepsilon}\int_{-\infty}^\infty \langle\delta\boldsymbol{x}(t)\delta\boldsymbol{x}^\intercal (0) \rangle_\text{ss} e^{-i\omega t} dt\,,
\end{align} 
where the scaling ensures that $\mathbb{Z}(\omega)$ remains finite in the limit $\varepsilon \rightarrow 0$.
A known result \cite{keizer2012statistical}, that we re-derive for completeness in \cref{sec:power}, is that
\begin{align}
\label{eq:PSD-omega}
    \mathbb{Z}(\omega) = (\mathbb{K}-i \omega \mathbb{1})^{-1} \mathbb{D} [(\mathbb{K}-i \omega \mathbb{1})^{-1}]^\dagger\,,
\end{align}
where $A^\dagger \equiv (\overline{A})^\intercal$ denotes the adjoint operation.
Two important and complementary quantities can be derived from it. 
First, the \textit{zero frequency PSD}, $\mathbb{Z}(0)$, which by time integrating the steady state correlation function, captures the persistence of fluctuations~\cite{berg1977physics}.
Second, the \textit{stationary covariance} 
which is obtained from it as $\mathbb{C}_\text{ss}=(2 \pi)^{-1}\int_{-\infty}^{\infty} d\omega \mathbb{Z}(\omega)$ and measures the magnitude and the directionality of fluctuations around the steady state. It can be calculated as the steady-state solution of the Lyapunov equation 
\begin{align}
\label{eq:Lyapunov}
   \mathbb{K} \mathbb{C}_\text{ss} +\mathbb{C}_\text{ss}  \mathbb{K}^\intercal = - \mathbb{D}\,.
\end{align}
Efficient ways to calculate \Cref{eq:PSD-omega,eq:Lyapunov} are well known~\cite{simpson2003frequency,warren2006exact,rodrigo2019ab,gupta2022frequency,mckane2007amplified,thomas2013signatures,adamer2020coloured,fyodorov2025nonorthogonal,song2019frequency}. 
Our goal is to relate them to the experimentally accessible physical responses. 

\textit{Nonequilibrium response---}We assume that the parameters $\boldsymbol{\theta}=(\theta_1,\dots,\theta_{N_p})$ control $\boldsymbol{f}(\boldsymbol{x},\boldsymbol{\theta})$ in the deterministic dynamics \cref{eq:dynamics-general} and that the system is initially at steady state, $\mathcal{X}(0)=\boldsymbol{x}^*(\boldsymbol{\theta})$.
We then consider a small perturbation of the model parameters, $\delta\boldsymbol{\theta}$, at $t=0$. 
Given that the system is stable, $\mathcal{X}(t)$ will eventually relax to the new steady state $\boldsymbol{x}^*(\boldsymbol{\theta} +\delta\boldsymbol{\theta})$. 
Since the perturbation is small, $\delta \mathcal{X}(t)=\mathcal{X}(t)-\boldsymbol{x}^*(\boldsymbol{\theta})$ will also be small and obeys the dynamics
\begin{subequations}
\label{eq:determ-dyn}
\begin{align}
\label{eq:determ-dyn-eq}
    d_t \delta \mathcal{X}(t) &= f\big(\boldsymbol{x}^*(\boldsymbol{\theta})+\delta \mathcal{X}(t),\boldsymbol{\theta}+\delta\boldsymbol{\theta}\big)\\
    &= \mathbb{K} \delta \mathcal{X}(t)+\mathbb{Q} \delta \boldsymbol{\theta} +\mathcal{O}(\delta\boldsymbol{\theta}^\intercal\delta \mathcal{X})\,,\label{eq:determ-ic}
\end{align}
\end{subequations}
where $\mathbb{K} \equiv \mathbb{K}(\boldsymbol{\theta})$ and 
\begin{align}
\label{eq:matQ}
 \mathbb{Q}
    \equiv \Big[ \frac{\partial f_n(\boldsymbol{x}^*(\boldsymbol{\theta}),\boldsymbol{\theta})}{\partial \theta_{k}} \Big]_{\{nk\}} 
    = - \mathbb{K} \frac{d\boldsymbol{x}^*(\boldsymbol{\theta})}{d \boldsymbol{\theta}}\,.
\end{align}
For the last equality, we expanded $f\big(\boldsymbol{x}^*(\boldsymbol{\theta}+\delta \boldsymbol{\theta}) ,\boldsymbol{\theta}+\delta\boldsymbol{\theta}\big)=0$ to first order in $\delta\boldsymbol{\theta}$. 
Combining the solution of \cref{eq:determ-ic} with \cref{eq:matQ}, the \textit{dynamical response} matrix can be written as
\begin{align}
\label{eq:response-matrix-time}
   \mathbb{R}(t) \equiv \Big[\frac{\delta \mathcal{X}_n(t)}{\delta\theta_k}\Big]_{\{nk\}} 
    = \mathbb{K}^{-1} (e^{\mathbb{K}t}-\mathbb{1}) \mathbb{Q}\;.
\end{align}
Since the fixed point of the dynamics is stable, all eigenvalues of $\mathbb{K}$, denoted $\lambda_n$, have negative real parts, $\text{Re}~\lambda_n < 0$, and thus $\mathbb{K}$ is invertible.
\begin{align}
\label{eq:matR-s}
    \hat{\mathbb{R}}(s) = -\frac{1}{s}(\mathbb{K}-s\mathbb{1})^{-1} \mathbb{Q} \,.
\end{align}
The \textit{static response} matrix is the $t \rightarrow \infty$ limit of the dynamical response, which, using \cref{eq:matR-s} and the final value theorem, can be expressed as
\begin{align}
    \label{eq:matR-static}
    \mathbb{R}_\text{ss}\equiv\mathbb{R}(\infty) = \lim_{s\to 0} s\hat{\mathbb{R}}(s) = - \mathbb{K}^{-1}\mathbb{Q}\,.
\end{align}

In a generic steady state, no relation is known between $\mathbb{R}_\text{ss}$ and $\mathbb{C}_\text{ss}$. However, in the special case of detailed balance dynamics (see \cref{sec:diff} for details), the stationary covariance becomes an equilibrium covariance $\mathbb{C}_\text{eq}$ satisfying $\mathbb{K}\mathbb{C}_\text{eq}=(\mathbb{K}\mathbb{C}_\text{eq})^\intercal=-\mathbb{D}/2$. The Lyapunov \cref{eq:Lyapunov} was used in the last equality. 
Therefore, inserting $\mathbb{K}^{-1}=-2\mathbb{C}_\text{eq}\mathbb{D}^{-1}$ into \cref{eq:matR-static}, we find the notorious relation between equilibrium response and equilibrium covariance
\begin{align}
    \label{eq:Kubo}
    \mathbb{R}_\text{eq} = 2\mathbb{C}_\text{eq}\mathbb{D}^{-1}\mathbb{Q}\,.
\end{align}   

Before proceeding, let us recall that due to linearity, the dynamics for the average $\langle \boldsymbol{x}(t)\rangle$ following from \cref{eq:Langevin-state} is identical to the deterministic dynamics of $\mathcal{X}(t)$ linearized around the fixed point. As a result, the present response theory is also a response theory for averages. Let us also stress that in experiments, averages (and thus their responses) and PSD are measurable quantities.

\textit{Linear dynamics from response---}We first note that $\mathbb{Q}$, using \cref{eq:response-matrix-time}, can be expressed in terms of the measurable dynamical response function
\begin{align} \label{eq:Jacobians-empirical}
\mathbb{Q}&=d_t \mathbb{R}(0) = \lim_{t\to 0}\frac{1}{t}\mathbb{R}(t)\,.
\end{align}
As a result, using \cref{eq:matR-static}, we find that $\mathbb{K}$ can be obtained from the measurable responses as
\begin{align} \label{eq:K-Response}
\mathbb{K}
=- d_t \mathbb{R}(0) \mathbb{R}_\text{ss}^{-1} 
= - \lim_{t\to 0}\frac{1}{t}\mathbb{R}(t)\mathbb{R}_\text{ss}^{-1}\,.
\end{align}
In doing so, we assumed that $\mathbb{Q}$ is invertible, which implies that the number of independent parameters $N_p$ is equal to or greater than the number of dynamical variables $N$. In the latter case, the Moore--Penrose inverse can be used. 

\textit{Linking response and PSD: Macroscopic FRRs---}Using \cref{eq:matR-s}, the resolvent $(\mathbb{K}-s\mathbb{1})^{-1}$ in \cref{eq:PSD-omega} can be rewritten as $(\mathbb{K}-s\mathbb{1})^{-1} = - s\hat{\mathbb{R}}(s)\mathbb{Q}^{-1}$. 
As a result, \cref{eq:PSD-omega} can be written as
\begin{align}
\label{eq:FRR-dynamic-matrix}
\mathbb{Z}(\omega) = \omega^2 \hat{\mathbb{R}}(i\omega)\cdot \mathbb{M} \cdot (\hat{\mathbb{R}}(i\omega))^\dagger\,,
\end{align}
where
\begin{align}
\label{eq:matM}
    \mathbb{M} = \mathbb{Q}^{-1}\mathbb{D}(\mathbb{Q}^{-1})^\intercal = \lim_{t\to 0}t^2\mathbb{R}^{-1}(t)\mathbb{D}(\mathbb{R}^{-1}(t))^\intercal\,.
\end{align}
Using \cref{eq:matR-static}, the zero frequency limit of \cref{eq:FRR-dynamic-matrix} reads
\begin{align}\label{eq:FRR-ss}
\mathbb{Z}(0) &= \mathbb{R}_\text{ss} \mathbb{M}\mathbb{R}_\text{ss}^\intercal \\
&= \lim_{t\to 0}t^2 \mathbb{R}(\infty) \mathbb{R}^{-1}(t)\mathbb{D}[\mathbb{R}(\infty) \mathbb{R}^{-1}(t)]^\intercal \,.
\end{align} 
The results [\cref{eq:FRR-dynamic-matrix,eq:matM,eq:FRR-ss}] are the macroscopic counterpart of the FRRs derived for Markov jump processes \cite{aslyamov2024frr, ptaszynski2024frr, ptaszynski2025frr-mix,bao2024nonequilibrium,aslyamov2025excess}. They demonstrate that FRRs preserve their structure for macroscopic dynamics in the weak noise limit.

\textit{Inferring the diffusion matrix---}Another important result, is that our approach provides three independent methods to determine the diffusion matrix that governs the stochastic dynamics of the system based on the measurable fluctuations and responses.
For the first method, we insert \cref{eq:K-Response} into the Lyapunov~\cref{eq:Lyapunov} and find
\begin{align}
\label{eq:matD-static}
\mathbb{D}&=d_t \mathbb{R}(0) \mathbb{R}_\text{ss}^{-1} \mathbb{C}_\text{ss}+\mathbb{C}_\text{ss} (\mathbb{R}_\text{ss}^{-1})^{\intercal} (d_t \mathbb{R}(0))^{\intercal} \nonumber\\
&=\lim_{t\to 0} \frac{1}{t}\Big[\mathbb{R}(t) \mathbb{R}_\text{ss}^{-1} \mathbb{C}_\text{ss}+\mathbb{C}_\text{ss} (\mathbb{R}_\text{ss}^{-1})^{\intercal} \mathbb{R}^{\intercal}(t)\Big]\,,
\end{align}
which infers the diffusion matrix in terms of static fluctuations and dynamic responses. 
For the second and third methods, we use the FRRs [\cref{eq:FRR-dynamic-matrix,eq:matM,eq:FRR-ss}] to isolate the diffusion matrix as 
\begin{subequations}
\label{eq:matD-PSD}
\begin{align}
\mathbb{D} &= \frac{1}{\omega^{2}} \mathbb{Q} \hat{\mathbb{R}}^{-1}(i\omega)\cdot \mathbb{Z}(\omega) \cdot (\hat{\mathbb{R}}^{-1} (i\omega))^\dagger \mathbb{Q}^{\intercal}\,,\\
\mathbb{D}&=\mathbb{Q} \mathbb{R}_\text{ss}^{-1} \cdot \mathbb{Z}(0)  \cdot (\mathbb{R}_\text{ss} ^{-1})^{\intercal} \mathbb{Q}^{\intercal} \,,
\end{align}
\end{subequations}
with $\mathbb{Q} = d_t \mathbb{R}(0)$. 

These results, together with \cref{eq:K-Response} confirm that $\mathbb{K}$ and $\mathbb{D}$ can be determined from the response of arbitrary steady state and its stationary fluctuations.

\textit{Response links static covariance and PSD---}We now show how the responses provide an explicit connection between static covariance and PSD.
Indeed, by multiplying Lyapunov~\cref{eq:Lyapunov} by $\mathbb{K}^{-1}$ from the left and $(\mathbb{K}^{-1})^\intercal$ from the right, then using \cref{eq:PSD-omega} for $\omega=0$, we get
\begin{align}
-\mathbb{C}_\text{ss} (\mathbb{K}^{-1})^\intercal - \mathbb{K}^{-1}\mathbb{C}_\text{ss} =\mathbb{K}^{-1} \mathbb{D} (\mathbb{K}^{-1})^\intercal=\mathbb{Z}(0) \,,
\end{align}
and using \cref{eq:Jacobians-empirical}, we find
\begin{align} \label{eq:covariance-PSD-relation}
\mathbb{C}_\text{ss} (\mathbb{R}_\text{ss} \mathbb{Q}^{-1})^{\intercal} + \mathbb{R}_\text{ss} \mathbb{Q}^{-1} \mathbb{C}_\text{ss} = \mathbb{Z}(0)\,, 
\end{align}
with $\mathbb{Q} = d_t \mathbb{R}(0)$. 
We note that \cref{eq:covariance-PSD-relation} has the same form as the Lyapunov equation \cref{eq:Lyapunov} and allows the static covariance to be expressed in terms of measurable quantities [\cref{sec:eq-cond}]. 
Moreover, we emphasize that \cref{eq:covariance-PSD-relation} holds as an equality only in the weak-noise limit. 
Indeed, in Appendix~\ref{app:schlogl} we use the Schl{\"o}gl model~\cite{schlogl1972chemical, vellela2009stochastic} (a Markov jump description of multimolecular chemical reactions) to show that the left-hand side of \cref{eq:covariance-PSD-relation} can be larger or smaller than its right-hand side and the equality only holds in the weak-noise limit. 
Therefore, \cref{eq:covariance-PSD-relation} can be used to assess whether the weak-noise approximation is valid in a given experiment.

Using \cref{eq:FRR-dynamic-matrix,eq:FRR-ss}, we can also relate the finite- and zero-frequency PSD through the responses:
\begin{align} 
&\mathbb{Z}(\omega) = \omega^2 \hat{\mathbb{R}}(i\omega)\cdot \mathbb{R}_\text{ss}^{-1} \mathbb{Z}(0) (\mathbb{R}_\text{ss}^{-1})^{\intercal} \cdot (\hat{\mathbb{R}}(i\omega))^\dagger\,.
\end{align}

\begin{figure}
    \centering
    \includegraphics[width=\linewidth]{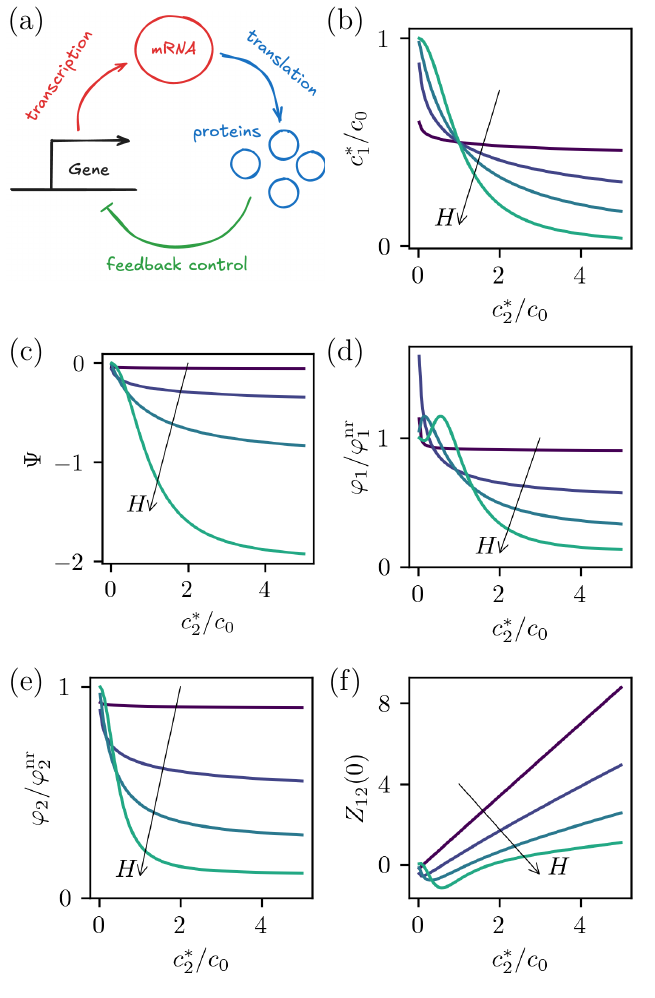}
    \caption{(a): Sketch of a simple transcription-translation process from gene to mRNAs to proteins with negative feedback control modeled by, $w_{+1}=k_m/[1+(\mathcal{C}_2/c_0)^H]$. 
    (b): fixed points $c^*_1$ and $c^*_2$; 
    (c): $\Psi(c^*_2)$ for the negative feedback;
    (d): mRNA scaled PSD $\varphi_1$;
    (e): protein scaled PSD $\varphi_2$; 
    (f): mRNA-proteins PSD covariance $Z_{12}(0)$. 
    Arrows denotes the direction of increasing parameter $H=0.1, 0.5, 1,2$. For calculations we used: $\tau_1=1$, $\tau_2 = 5$, $c_0 = 1$, $k_m = 1$. 
    }
    \label{fig:fig-2}
    \vspace{-0.5cm}
\end{figure}

\textit{Noise in gene regulatory networks---}Genetically identical cells exhibit significant variability in their molecular composition and behavior under uniform conditions due to random fluctuations in gene expression. 
We derive analytical expressions for the PSDs of mRNA molecular numbers transcribed from a gene and the protein molecular numbers translated from mRNAs~\cite{kaern2005stochasticity}; see \cref{fig:fig-2}(a). 
The deterministic rate equations for the concentrations of mRNAs, $\mathcal{C}_1$, and proteins, $\mathcal{C}_2$, are
\begin{align}
\label{eq:genes-ode}
    d_t \mathcal{C}_1 &= f_1(\mathcal{C}_1,\mathcal{C}_2)\,,\quad d_t \mathcal{C}_2 = f_2(\mathcal{C}_1, \mathcal{C}_2)\,,
\end{align} 
with $f_n = w_{+ n} - w_{-n}$ and
\begin{align}
\label{eq:genes-model-linear}
w_{+1} &=  w_{+1}(\mathcal{C}_2)\,,\,  w_{-1} = \frac{\mathcal{C}_1}{\tau_1}\,,\,w_{+2} = k_p \mathcal{C}_1\,,\, w_{-2} = \frac{\mathcal{C}_2}{\tau_2}\,,
\end{align}
where the mRNA synthesis rate $w_{+1}$ is an arbitrary function of $\mathcal{C}_2$, $\tau_1$ and $\tau_2$ are the degradation time scales of mRNA and protein, respectively, and $k_p$ is the protein synthesis rate. 
The fixed point is $c^*_n = w_{+n}\tau_n$. 

Our macroscopic fluctuation theory holds when the volume $\Omega = 1/\varepsilon$ is large in \cref{eq:Langevin-state} \cite{van1992stochastic, falasco2023macroscopic}. The PSD for the concentrations thus reads $Z_{mn}(0) = \Omega \int_{-\infty}^\infty \langle \delta c_m(t)\delta c_n(0) \rangle_\text{ss}dt$, where $\delta c_n = c_n - c_n^*$. 
For analytical calculations, we can always use perturbation parameters $\boldsymbol{\theta}$ such that $\mathbb{Q}=1$ and $\mathbb{M} = \mathbb{D}$. Indeed, if we perturb the rates, $\boldsymbol{\theta} = (w_{+1}, w_{+2})^\intercal$, then $\mathbb{Q}=\mathbb{1}$ and $\mathbb{M} = \text{diag}(D_1, D_2)$, where $D_1 = 2w_{\pm 1}$ and $D_2 = 2w_{\pm 2}$ are the diffusion coefficients~\cite{paulsson2004summing,paulsson2005models,bruggeman2009noise}, and the FRR [\cref{eq:FRR-ss}] read
\begin{align}
\label{eq:FRR-genes}
     Z_{mn}(0) =  \sum_{k=1}^2 D_k \frac{d c_m^*}{dw_{+k}} \frac{d c_n^*}{d w_{+k}}\,,
\end{align}
where the right-hand side is evaluated at the fixed point. 
To proceed with \cref{eq:FRR-genes}, one needs the static responses
\begin{align}
\label{eq:responses-genes}
    \frac{d c^*_n}{d w_{+n}} &= \frac{\tau_m^{-1}}{\det \mathbb{K}}\,,\quad 
    \frac{d c^*_n}{d w_{+m}} = \frac{1}{\det \mathbb{K}} \frac{\partial w_{+n}}{\partial \mathcal{C}_m}\,,
\end{align}
where the derivative $\partial \mathcal{C}_m$ are calculated at the fixed point and with the determinant $\det \mathbb{K}$ calculated from \cref{eq:genes-ode} as
\begin{subequations}
\label{eq:detK-Psi}
\begin{align}
\label{eq:detK}
    \det K &= \frac{1}{\tau_1 \tau_2} - \frac{\partial w_{+1}}{\partial \mathcal{C}_2}\frac{\partial w_{+2}}{\partial \mathcal{C}_1}= \frac{1}{\tau_1 \tau_2}(1-\Psi)\,,\\
    \label{eq:Psi}
\Psi &\equiv \frac{\partial \ln (w_{+1}/w_{-1})}{\partial \ln \mathcal{C}_2} =  \frac{\partial\ln w_{+1}}{\partial \ln \mathcal{C}_2}\,,
\end{align}
\end{subequations}
where $\Psi$ is the logarithmic gain defined at the fixed point (it corresponds to $H_{12}$ in \cite{paulsson2004summing}). 
We note that $\Psi \leq 1$ for a stable fixed point and $\Psi \leq 0$ for a negative feedback. 
In simulations, we model the negative feedback of protein molecules on mRNA synthesis by the function $w_{+1}=k_m/[1+(\mathcal{C}_2/c_0)^H]$, where $k_m$ is the rate constant, $c_0$ is the scale parameter and $H$ is the power of the feedback. This implies the Hill function $\Psi_\text{Hill} = - H(c^*_2/c_0)^H/[1+ (c^*_2/c_0)^H]$, which satisfies $-H\leq\Psi_\text{Hill}\leq 0$; see \cref{fig:fig-2}(c). In absence of feedback, $H=0$. 

Turning to fluctuations, the mRNA fluctuations, using \cref{eq:FRR-genes,eq:responses-genes,eq:detK-Psi}, are described by the scaled PSD
\begin{align}
\label{eq:eps-1-feedback}
    \varphi_1 \equiv \frac{Z_{11}(0)}{c^*_1\tau_1} = \frac{2}{(1-\Psi)^2}\Big[1 + \frac{\tau_2}{\tau_1} \frac{c^*_1}{c^*_2}\Psi^2\Big]= \frac{2[1+\Psi^2(\tau_1 k_p)^{-1}]}{(1-\Psi)^2}\,,
\end{align}
where we used $c^*_2 = k_p c^*_1 \tau_2$ implying $\tau_2 c^*_1/(\tau_1 c^*_2) = (\tau_1 k_p)^{-1}$. 
In the absence of feedback, $\Psi = 0$, we find $\varphi_1^\text{nr}=2$. 
For negative feedbacks, the scaled $\varphi_1$ from \cref{eq:eps-1-feedback} can be higher or lower than $\varphi^\text{nr}_1$; see \cref{fig:fig-2}(d). 
In turn, protein fluctuations, calculated using \cref{eq:FRR-genes,eq:responses-genes,eq:detK-Psi}, are described by
\begin{align}
\label{eq:eps-2-feedback}
\varphi_2 &\equiv \frac{Z_{22}(0)}{c_2^*\tau_2}= \frac{2}{(1-\Psi)^2}\Big[1+ \frac{\tau_1}{\tau_2}\frac{c^*_2}{c^*_1} \Big] = \frac{1}{(1-\Psi)^2} \varphi_2^\text{nr}\,,
\end{align}
where the non-regulated value is
\begin{align}
\label{eq:protein-var-result}
     \varphi^\text{nr}_2 & =2\Big[ 1 + \frac{\tau_1}{\tau_2}\frac{c^*_2}{c^*_1}\Big] = 2(1+\tau_1 k_p)\,,
\end{align}
which predicts that mRNA with a longer life span induce a stronger protein noise while decreasing mRNA noise [see \cref{eq:eps-1-feedback}]. 
From \cref{eq:eps-2-feedback}, we see that negative feedback, $\Psi < 0$, suppresses protein fluctuations: $\varphi_2 \leq \varphi_2^\text{nr}$. Thus, between two identical setups, the one with negative feedback produces weaker noise (zero-frequency PSD); see \cref{fig:fig-2}(e).
Finally, cross-correlations between mRNAs and proteins read
\begin{align}
    \label{eq:gene-covariance}
    Z_{12}(0) & = D_1 \frac{d c^*_1}{d w_{+1}}\frac{d c^*_2}{d w_{+1}} + D_2 \frac{d c^*_1}{d w_{+2}}\frac{d c^*_2}{d w_{+2}} \nonumber \\
    &=\frac{2}{(1-\Psi)^2}(\tau_1 c^*_2 + \tau_2 c^*_1 \Psi) \,.
\end{align}
This shows that a change of sign in $Z_{12}(0)$ necessarily reveals the presence of negative feedbacks, $\Psi < 0$, see \cref{fig:fig-2}(f).

Importantly, our theory can be applied to arbitrary complex gene regulatory networks, i.e. networks described by $d_t\boldsymbol{C} = \boldsymbol{w}_{+}(\boldsymbol{C})-\text{diag}(\boldsymbol{\tau})^{-1}\boldsymbol{C}$, where $\boldsymbol{C}$ is the vector of concentrations, $\boldsymbol{w}_{+}$ an arbitrary vector dependent on $\boldsymbol{C}$, and $\text{diag}(\boldsymbol{\tau})$ the diagonal matrix of degradation time scales. In such a case, \cref{eq:FRR-ss} can be written as
\begin{align}
\label{eq:FRR-int-ext}
    Z_{nn}(0) = \underbrace{D_{n}\Big(\frac{d c_n^*}{dw_{+n}}\Big)^2}_{\text{int.}\geq 0}+\underbrace{\sum_{k\neq n} D_{k} \Big(\frac{d c_n^*}{dw_{+k}}\Big)^2}_{\text{ext.}\geq 0}\,,
\end{align}
which reveals a decomposition of the noise (PSD) into intrinsic and extrinsic contributions, distinguished by local ($k=n$) and nonlocal ($k \neq n$) response terms. The ``intrinsic'' and ``extrinsic'' terminology is common in the literature~\cite{elowitz2002stochastic}, but finds a clear mathematical formulation at the level of the PSD within our theory.
Regardless of how silent the extrinsic network may be (when its responses are negligible), the intrinsic PSD in \cref{eq:FRR-int-ext} is nonzero. As in the case of stationary fluctuations~\cite{elowitz2002stochastic}, the intrinsic PSD sets the fundamental lower limit on fluctuations in complex gene networks. 

\textit{Conclusions---}We presented a linear nonequilibrium fluctuation-response theory describing macroscopic systems close to fixed stable points. Every ingredient of the theory can be reconstructed from the experimentally accessible response and PSD. We also illustrated the analytical potential of the theory by considering fluctuations in gene regulatory networks.  
Extending the theory to more complex attractors such as limit cycles~\cite{santolin2025dissipation,nagayama2025duality} is left as a future perspective. 

\textit{Author’s note---}On this day of submission, \cite{dechant2025finite} was posted on arXiv. This work contains overlaps with ours. It shows that our \cref{eq:matD-PSD} holds as an inequality for nonlinear Langevin equations, including underdamped ones.

\begin{acknowledgments}
T.A. and M.E. are funded by the Fonds National de la Recherche-FNR, Luxembourg: project ThermoElectroChem (C23/MS/18060819) and NEQPHASETRANS (C24/MS/18933049), respectively.
K.P. is funded by the National Science Centre, Poland: project No.\ 2023/51/D/ST3/01203.  
\end{acknowledgments}

\begin{center}
  \large \bf End Matter
\end{center}
\vspace{-1cm}
\appendix

\section{Deriviation of \cref{eq:PSD-omega}}
\label{sec:power}
We first formally solve \cref{eq:Langevin-state} for a single noise realization starting in $\boldsymbol{x}^*$ at $t \rightarrow -\infty$
\begin{align}
\label{eq:state-sol-general}
\delta\boldsymbol{x}(t) 
=\sqrt{\varepsilon} \int_{-\infty}^t e^{\mathbb{K}(t-t')}\boldsymbol{\eta}(t') dt'\,,
\end{align}
and use it to prove that
\begin{align}
\label{eq:corr-func-t}
\langle \delta \boldsymbol{x}(t) \delta \boldsymbol{x}^\intercal(0) \rangle_\text{ss} =e^{\mathbb{K}t} \langle  \delta \boldsymbol{x}(0) \delta \boldsymbol{x}^\intercal(0) \rangle_\text{ss}  = e^{\mathbb{K}t} \mathbb{C}_\text{ss}\,.
\end{align}
Then from the time-translation symmetry of stationary correlation functions, we get
\begin{align}
    \langle \delta \boldsymbol{x}(-t) \delta \boldsymbol{x}^\intercal(0) \rangle_\text{ss} 
    =\langle \delta \boldsymbol{x}(0) \delta \boldsymbol{x}^\intercal(t) \rangle_\text{ss} 
    =\langle \delta \boldsymbol{x}(t) \delta \boldsymbol{x}^\intercal(0) \rangle^\intercal_\text{ss}  \,,
\end{align}
where \cref{eq:corr-func-t} was used for the last equality.
Using it in \cref{eq:PSD}, we find
\begin{align}
    \mathbb{Z}(\omega) & = \int_0^\infty dt\Big(e^{t(\mathbb{K}-i \omega \mathbb{1})} \mathbb{C}_\text{ss} + \mathbb{C}_\text{ss} e^{t(\mathbb{K}-i \omega \mathbb{1})^\dagger} \Big)\nonumber\\ 
    &= 
    - (\mathbb{K}-i \omega \mathbb{1})^{-1} \mathbb{C}_\text{ss} - \mathbb{C}_\text{ss}  \,[(\mathbb{K}-i \omega \mathbb{1})^{-1}]^\dagger\nonumber\\
    &=-(\mathbb{K}-i \omega \mathbb{1})^{-1}\{\mathbb{C}_\text{ss}(\mathbb{K}-i \omega \mathbb{1})^\dagger\nonumber\\
    &\; \; \; \;+(\mathbb{K}-i \omega \mathbb{1})\mathbb{C}_\text{ss}\}[(\mathbb{K}-i \omega \mathbb{1})^{-1}]^\dagger \nonumber\\
    &= (\mathbb{K}-i \omega \mathbb{1})^{-1} \mathbb{D} [(\mathbb{K}-i \omega \mathbb{1})^{-1}]^\dagger\,,
\end{align}
where we use the Lyapunov equation \cref{eq:Lyapunov} for the last equality. 

\section{Fokker--Plank equation}
\label{sec:diff}
The Fokker--Plank equation associated to the Langevin dynamics \cref{eq:Langevin-state} reads
\begin{subequations}
\begin{align}
    \label{eq:FP}
    \partial_t P(\boldsymbol{x},t) &= 
    -\nabla_{\boldsymbol{x}}^\intercal \big[\boldsymbol{v}(\boldsymbol{x},t) P(\boldsymbol{x}, t)\big]\,,\\
    \label{eq:FP-curent}
    v(\boldsymbol{x},t) &= \mathbb{K}(\boldsymbol{x} - \boldsymbol{x}^*)  + \frac{1}{2} \mathbb{D} \nabla_{\boldsymbol{x}} \log P(\boldsymbol{x},t)\,,
\end{align}
\end{subequations}
where $\boldsymbol{v}$ is the probability velocity and $\nabla_{\boldsymbol{x}}=(\dots,\partial_{x_n},\dots)^\intercal$. 
Due to its linearity, its solution is the Gaussian
\begin{align}
\label{eq:Gauss-dydamics}
    P = \frac{\exp\big[-\tfrac{1}{2}\delta\boldsymbol{x}^\intercal (t)\mathbb{C}^{-1} \delta\boldsymbol{x}(t) \big]}{\sqrt{(2\pi)^N\det\mathbb{C}}}\,,
\end{align}
where the dynamical covariance, $\mathbb{C}(t)$, satisfies the dynamical Lyapunov equation
\begin{align}
    \label{eq:Lyapunov-dynamics}
    d_t \mathbb{C} = \mathbb{K} \mathbb{C} +\mathbb{C}  \mathbb{K}^\intercal + \mathbb{D}\,,
\end{align}
with initial condition $\mathbb{C}(0) = 0$. 
At steady-state
\begin{align}
\label{eq:Gauss-ss}
    P_\text{ss} = \frac{\exp\big[-\tfrac{1}{2}(\boldsymbol{x}-\boldsymbol{x}^*)^\intercal\mathbb{C}_\text{ss}^{-1}(\boldsymbol{x}-\boldsymbol{x}^*)\big]}{\sqrt{(2\pi)^N\det\mathbb{C}_\text{ss}}}\,,
\end{align}
and velocity
\begin{align}    
   \boldsymbol{v}_\text{ss} &=  
   \big(\mathbb{K} + \tfrac{1}{2}\mathbb{D}\mathbb{C}_\text{ss}^{-1}\big)(\boldsymbol{x}-\boldsymbol{x}^*)=\mathbb{V}_\text{ss}(\boldsymbol{x}-\boldsymbol{x}^*)\,.
\end{align}
The dynamics is said to be detailed balance when
$\boldsymbol{v}_\text{eq} = 0$ which implies
\begin{align}
\label{LDBCond}
\mathbb{K} = - \tfrac{1}{2}\mathbb{D}\mathbb{C}_\text{eq}^{-1}\;.
\end{align}
This condition is equivalent to requesting the time-reversibility of the correlation functions
\begin{align}
\label{eq:time-reverse}
    \langle\delta\boldsymbol{x}(t)\delta\boldsymbol{x}^\intercal (0) \rangle_\text{ss}
    =\langle\delta\boldsymbol{x}(-t)\delta\boldsymbol{x}^\intercal (0) \rangle_\text{ss}
    =\langle\delta\boldsymbol{x}(0)\delta\boldsymbol{x}^\intercal (t) \rangle_\text{ss}\,,
\end{align}
where the last equality follows from time-translation invariance at steady state. 
Indeed, using \cref{eq:corr-func-t}, \cref{eq:time-reverse} can be rewritten as 
\begin{align}
\label{eq:time-reverse-cond}
    e^{\mathbb{K}t}\mathbb{C}_\text{ss} = \mathbb{C}_\text{ss}e^{\mathbb{K}^\intercal t} \quad\rightarrow \quad \mathbb{K}\mathbb{C}_\text{ss} = \mathbb{C}_\text{ss} \mathbb{K}^\intercal \,,
\end{align}
which, together with the stationary Lyapunov equation, means that $\mathbb{C}_\text{ss}=\mathbb{C}_\text{eq}$ in \cref{LDBCond}. 

\section{Sufficient time-reversibility condition}
\label{sec:eq-cond}
Using Eq.~(5.19) in \cite{lax1960fluctuations}, the Lyapunov \cref{eq:Lyapunov} is solved by
\begin{align}
    \mathbb{C}_\text{ss} = -\int_0^\infty e^{\mathbb{K}t}\mathbb{D}e^{\mathbb{K}^\intercal t} dt \,.
\end{align}
When inserted in the time-reversibility condition \eqref{eq:time-reverse-cond}, we get
\begin{align}
\label{eq:sufficient-cond-1}
    \mathbb{K}\mathbb{C}_\text{ss} - \mathbb{C}_\text{ss} \mathbb{K}^\intercal = -\int_0^\infty e^{\mathbb{K}t}(\mathbb{K}\mathbb{D}-\mathbb{D}\mathbb{K}^\intercal )e^{\mathbb{K}^\intercal t} dt\,.
\end{align}
Thus, the symmetry $(\mathbb{K}\mathbb{D})^\intercal = \mathbb{K}\mathbb{D}$ implies time-reversibility. 

Similarly, we can solve \cref{eq:covariance-PSD-relation} as
\begin{align}
    \mathbb{C}_\text{ss} = - \int_0^\infty e^{-\mathbb{R}\mathbb{Q}^{-1}t}\mathbb{Z}(0)e^{-[\mathbb{R}\mathbb{Q}^{-1}]^\intercal t} dt \,,
\end{align}
where $\mathbb{R}\mathbb{Q}^{-1} = - \mathbb{K}^{-1}$. Since $\mathbb{K}$ and $e^{-\mathbb{K}^{-1}}$ commute, a sufficient condition for time-reversibility, \cref{eq:time-reverse-cond,eq:sufficient-cond-1}, is also
\begin{align}
\label{eq:sufficient-cond-2}
    d_t \mathbb{R}_\text{eq}(0)\mathbb{R}_\text{eq}^{-1}\mathbb{Z}_\text{eq}(0) - \mathbb{Z}_\text{eq}(0)[d_t \mathbb{R}_\text{eq}(0)\mathbb{R}_\text{eq}^{-1}]^\intercal = 0\,,
\end{align}
where we used \cref{eq:Jacobians-empirical}.
Unlike \cref{eq:sufficient-cond-1}, \cref{eq:sufficient-cond-2} is only expressed in terms of measurable quantities. 

\section{Asymptotic validity of \cref{eq:covariance-PSD-relation}} \label{app:schlogl}

Our goal is to test the identity \eqref{eq:covariance-PSD-relation} in a model that can operate inside or outside of the weak noise regime.
We consider the Schl{\"o}gl model~\cite{schlogl1972chemical,vellela2009stochastic,gaspard2004fluctuation}, a paradigmatic model of nonlinear chemical kinetics. The chemical reactions
\begin{align}
\ce{A + 2X <=>[k_{+1}][k_{-1}] 3X} \,, \quad
\ce{B <=>[k_{+2}][k_{-2}] X} \,,
\end{align}
occur in a well-mixed container with volume $\Omega$. The concentrations of species A and B, denoted $c_A$ and $c_B$, are kept constant (chemostatted), while the number of molecules X, denoted by $n$, fluctuates. 
The dynamics for $n$ is a Markov jump process and the probability of $n$, $\pi_n$, evolves according to the chemical master equation
\begin{align}
d_t \boldsymbol{\pi}= \mathbb{W} \boldsymbol{\pi} \,,
\end{align}
where the rate matrix $\mathbb{W}$ has elements 
\begin{align}
W_{nm}=\delta_{m,n-1} g_{n-1}^++\delta_{m,n+1} g_{n+1}^--\delta _{nm} (g_n^++g_n^-) \,,
\end{align}
with
\begin{subequations}
\begin{align}
g_n^+ &=[k_{+1} c_A n(n-1)/ \Omega +k_{+2} c_B \Omega](1-\delta_{nn_c}) \,, \\
g_n^- &=k_{-1} n(n-1)(n-2)/ \Omega^{2} +k_{-2} n \,.
\end{align}
\end{subequations}
In the numerical treatment, we truncate the molecule number $n_c \geq n,m \geq 0$.
The concentration of species X is defined as $x=n/\Omega$. 
In the large volume limit, $\Omega \to \infty$, the probability distribution for $x$ concentrates around its most likely value $\mathcal{X}(t)$, which obeys the deterministic rate equation
\begin{align}
d_t \mathcal{X}=f(\mathcal{X})=w_+(\mathcal{X})-w_-(\mathcal{X}) \,,
\end{align}
with $w_+(x)=k_{+1} c_A x^2+k_{+2} c_B$ and $w_-(x)=k_{-1} x^3+k_{-2} x$. 
We focus on the regime where the system has a unique stable fixed point $f(x^*)=0$. 
In the large but not infinite volume limit, the dynamics around the fixed point is described by the linear Langevin dynamics \cref{eq:dynamics-general}, where $\varepsilon=\Omega^{-1}$, $K=\partial_x f(x)|_{x=x^*}$ and $D=w_+(x^*)+w_-(x^*)$. In this regime, $\langle x(t) \rangle = \mathcal{X}(t)$ and
\cref{eq:covariance-PSD-relation} reads
\begin{align} \label{eq:mathcal-r}
\mathcal{R} \equiv \frac{Z(0) d_{c_A} d_t \langle x(0) \rangle}{2 C_\text{ss} d_{c_A} \langle x(\infty) \rangle}=1 \,,
\end{align}
where $C_\text{ss}=\Omega \langle \delta x^2(t) \rangle_\text{ss}$ and $Z(0) = \Omega\int_{-\infty}^\infty \langle \delta x(t)\delta x(0) \rangle_\text{ss}dt$. We choose to perturb the parameter $\theta=c_A$. 
Since $\mathcal{R}$ is only expressed in terms of empirical quantities, we want to test whether \cref{eq:mathcal-r} also holds outside of the weak noise limit.
To do so, we use standard master equation methods to compute responses and fluctuations~\cite{ptaszynski2024critical, ptaszynski2024frr, lapolla2020spectral, lapolla2019manifestations, lapolla2018unfolding}, and find
\begin{subequations}
\begin{align}
&d_{c_A} \langle x(\infty) \rangle = - \boldsymbol{n}^\intercal \mathbb{W}^D (d_{c_A} \mathbb{W}) \boldsymbol{\pi}_\text{ss}/\Omega \,, \\
&d_{c_A} d_t \langle x(0) \rangle = \boldsymbol{n}^\intercal  (d_{c_A} \mathbb{W}) \boldsymbol{\pi}_\text{ss}/\Omega \,, \\
&C_\text{ss} =\left[ \boldsymbol{n}^\intercal \cdot \text{diag}(\boldsymbol{\pi}_\text{ss}) \cdot \boldsymbol{n}-\left( \boldsymbol{n}^\intercal \boldsymbol{\pi}_\text{ss} \right)^2 \right]/\Omega\,, \\
&Z(0) = \\ \nonumber
&\hspace{0.3cm} -\boldsymbol{n}^\intercal \cdot \Big\{ \mathbb{W}^D \cdot \text{diag}(\boldsymbol{\pi}_\text{ss}) + \big[\mathbb{W}^D \cdot\text{diag}(\boldsymbol{\pi}_\text{ss})  \big]^\intercal  \Big\} \cdot \boldsymbol{n}/\Omega \,,
\end{align}
\end{subequations}
where $\boldsymbol{\pi}_\text{ss}$ is the stationary probability vector $\mathbb{W} \boldsymbol{\pi}_\text{ss}=0$, $\boldsymbol{n}=(0,1,\ldots,n_c)^\intercal$, and $\mathbb{W}^D$ is the Drazin inverse of the matrix $\mathbb{W}$.
\begin{figure}
    \centering
    \includegraphics[width=0.7\linewidth]{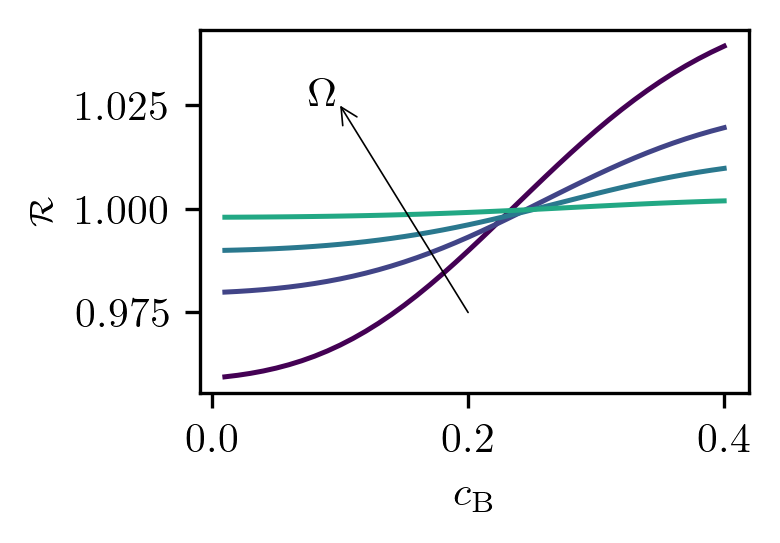}
    \vspace{-0.35cm}
    \caption{The parameter $\mathcal{R}$ defined in \cref{eq:mathcal-r} as a function of $c_B$. The arrow denotes direction of the increasing $\Omega=50,100,200,1000$. Parameters: $c_A=k_{\pm 1}=k_{\pm 2}=1$, $n_c=2\Omega$.
    }  
    \label{fig:app-schlogl}
\end{figure}
Plotting the ratio $\mathcal{R}$ for different volumes $\Omega$ in Fig.~\ref{fig:app-schlogl}, we observe that $\mathcal{R}$ converges to 1 only when $\Omega \to \infty$. This shows that \cref{eq:mathcal-r}, and thus \cref{eq:covariance-PSD-relation}, are only valid in the weak noise regime.

\bibliography{biblio}
\end{document}